# Low Noise, tunable Ho:fiber Soliton Oscillator for Ho:YLF Amplifier Seeding

**Peng Li[1], Axel Ruehl[1,*], Colleen Bransley[1,2] and Ingmar Hartl[1]**

[1]*Deutsches Elektronen-Synchrotron (DESY), Notkestrasse 85, 22607 Hamburg, Germany*
[2]*Physics Department, University of Dayton, 300 College Park, Dayton, OH 45469, United States*
[*]*axel.ruehl@desy.de*

**Abstract:** We present a passively mode-locked, tunable soliton Ho:fiber ring oscillator, optimized for seeding of Ho:YLF amplifiers. The oscillator is independently tunable in central wavelength and spectral width from 2040 nm to 2070 nm and from 5 nm to 10 nm, respectively. At all settings the pulse energy within the soliton is around 800 pJ. The soliton oscillator was optimized to fully meets the spectral requirements for seeding Ho:YLF amplifiers. Its Kelly sidebands are located outside the amplifier gain spectrum, resulting in a train of about 1 ps long pedestal-free pulses with relative intensity noise (RIN) of only 0.13 % RMS when integrated from 1 Hz to Nyquist frequency.

**References and links**

1. K. Scholle, S. Lamrini, P. P. Koopmann, and P. Fuhrberg, in Frontiers in Guided Wave Opt. and Optoelectronics, B. Pal, ed. (InTech, 2010), pp. 471–500.
2. N. Leindecker, A. Marandi, R.L. Byer, K.L. Vodopyanov, J. Jiang, I. Hartl, M.E. Fermann, and P.G. Schunemann, "Octave-spanning ultrafast OPO with 2.6-6.1µm instantaneous bandwidth pumped by femtosecond Tm-fiber laser," Opt. Express **20**, 7046-7053 (2012).
3. C.R. Phillips, J. Jiang, C. Mohr, A.C. Lin, C. Langrock, M. Snure, D. Bliss, M. Zhu, I. Hartl, J.S. Harris, M.E. Fermann, and M.M. Fejer, "Widely tunable mid-infrared difference frequency generation in orientation-patterned GaAs pumped with a femtosecond Tm-fiber system," Opt. Lett. **37**, 2928-2930 (2012).
4. G. Imeshev, M. E. Fermann, K. L. Vodopyanov, M. M. Fejer, X. Yu, J. S. Harris, D. Bliss, and C. Lynch, "High-power source of THz radiation based on orientation-patternd GaAs pumped by a fiber laser," Opt. Express **14**, 4439-4444 (2006).
5. P. Malevich G. Andriukaitis, T. Flöry, A. J. Verhoef, A. Fernández, S. Ališauskas, A. Pugžlys, A. Baltuška, L.H. Tan, C.F. Chua, and P.B. Phua, "High energy and average power femtosecond laser for driving mid-infrared optical parametric amplifiers," Opt. Lett. **38**, 2746-2748 (2013).
6. M. Hemmer, D. Sanchez, M. Jelínek, H. Jelínková, V. Kubeček, and J. Biegert, "Fiber-seeded, 10-ps, 2050-nm, multi-mJ, cryogenic Ho:YLF CPA," in CLEO: 2014, paper SM1F.3.
7. K. Murari, H. Cankaya, P. Li, A. Ruehl, I. Hartl, and F. X. Kärtner, "1.2 mJ, 1 kHz, ps-pulses at 2.05 µm from a Ho:fibre/Ho:YLF laser", in Europhoton Conference (2014), paper ThD-T1-O-05.
8. F. Haxsen, A. Ruehl, M. Engelbrecht, D. Wandt, U. Morgner, and D. Kracht, "Stretched-pulse operation of a thulium-doped fiber laser," Opt. Express **16**, 20471–20476 (2008)
9. K. Kieu and F.W. Wise, "Soliton thulium-doped fiber laser with carbon nanotube saturable absorber," IEEE Photon. Technol. Lett. **21**, 128–130 (2009).
10. S. Kivistö, T. Hakulinen, A. Kaskela, B. Aitchison, D.P. Brown, A.G. Nasibulin, E.I. Kauppinen, A.Härkönen, and O.G. Okhotnikov, "Carbon nanotube films for ultrafast broadband technology," Opt. Express **17**, 2358–2363 (2009).
11. J. Jiang, C. Mohr, J. Bethge, A. Mills, W. Mefford, I. Hartl, M. E. Fermann, C.-C. Lee, S. Suzuki, T. R. Schibli, N. Leindecker, K. L. Vodopyanov, and P. G. Schunemann, "500 MHz, 58 fs highly coherent Tm fiber soliton laser," in CLEO 2012, paper CTh5D.7.
12. F. Haxsen, D. Wandt, U. Morgner, J. Neumann, and D. Kracht, "Monotonically chirped pulse evolution in an ultrashort pulse thulium-doped fiber laser," Opt. Lett. **37**, 1014–1016 (2012).
13. A. Wienke, F. Haxsen, D. Wandt, U. Morgner, J. Neumann, and D. Kracht, "Ultrafast, stretched-pulse thulium-doped fiber laser with a fiber-based dispersion management," Opt. Lett. **37**, 2466–2468 (2012).

14. Y. Nomura, and T. Fuji, "Sub-50-fs pulse generation from thulium-doped ZBLAN fiber laser oscillator," Opt. Express **22**, 12461-12466 (2014).
15. Q. Wang, J. Geng, Z. Jiang, T. Luo, and S. Jiang, "Mode-Locked Tm-Ho-Codoped Fiber Laser at 2.06 μm," IEEE Photon. Technol. Lett. **23**, 682-684 (2011).
16. S. Kivistö, T. Hakulinen, M. Guina, and O. G. Okhotnikov, "Tunable Raman Soliton Source Using Mode-Locked Tm–Ho Fiber Laser," IEEE Photon. Technol. Lett. **19**. 934-936 (2007).
17. M. Yu. Koptev, E. A. Anashkina, A. V. Andrianov, S. V. Muravyev, and A. V. Kim, "Two-color optically synchronized ultrashort pulses from a TmYb-co-doped fiber amplifier," Opt. Lett. **39**, 2008-2010 (2014).
18. N. Nishizawa and T. Goto, "Widely wavelength-tunable ultrashort pulse generation using polarization maintaining optical fibers," IEEE J. Sel. Topics in Quantum Electron. **7**, 518-524 (2001).
19. G. Imeshev and M. Fermann, "230-kW peak power femtosecond pulses from a high power tunable source based on amplification in Tm-doped fiber," Opt. Express **13**, 7424-7431 (2005).
20. S. Kumkar, G. Krauss, M. Wunram, D. Fehrenbacher, U. Demirbas, D. Brida, and A. Leitenstorfer, " Femtosecond coherent seeding of a broadband Tm:fiber amplifier by an Erfiber system," Opt. Lett. **37**, 554-556 (2012).
21. F. Adler and S. A. Diddams, "High-power, hybrid Er:fiber/Tm:fiber frequency comb source in the 2 μm wavelength region," Opt. Lett. **37**, 1400-1402 (2012).
22. H. Hoogland, A. Thai, D. Sánchez, S. L. Cousin, M. Hemmer, M. Engelbrecht, J. Biegert, and R. Holzwarth, "All-PM coherent 2.05 μm Thulium/Holmium fiber frequency comb source at 100 MHz with up to 0.5 W average power and pulse duration down to 135 fs," Opt. Express **21**, 31390-31394 (2013).
23. N. Coluccelli, M. Cassinerio, A. Gambetta, P. Laporta, and G. Galzerano, "High-power frequency comb in the range of 2-2.15 um based on a holmium fiber amplifier seeded by wavelength-shifted raman solitons from an erbium-fiber laser," Opt. Lett. **39**, 1661-1663 (2014).
24. A. Chamorovsky A.V. Marakulin, S. Ranta, M. Tavast, J. Rautiainen, T. Leinonen, A.S. Kurkov, and O.G. Okhotnikov, "Femtosecond mode-locked holmium fiber laser pumped by semiconductor disk laser," Opt. Lett. **37**, 1448-1450 (2012).
25. A. Chamorovsky A.V. Marakulin, A.S. Kurkov, and O.G. Okhotnikov, "Tunable Ho-doped soliton fiber laser mode-locked by carbon nanotube saturable absorbers," Laser Phy. Lett. **9**, 602-606 (2012).
26. N. Coluccelli, A. Lagatsky, A. Di Lieto, M. Tonelli, G. Galzerano, W. Sibbeth, and P. Laporta, "Passive mode locking of an in-band-pumped Ho:YLiF4 laser at 2.06 μm," Opt. Lett. **36**, 3209-3211 (2011).
27. P. Li, A. Ruehl, U. Grosse-Wortmann, and I. Hartl, "Sub-100 fs passively mode-locked holmium-doped fiber oscillator operating at 2.06 μm," Opt. Lett. **39**, 6859-6862 (2014).
28. A. Ruehl, D. Wandt, U. Morgner, and D. Kracht, "On wave-breaking free fiber lasers mode-locked with two saturable absorber mechanisms," Opt. Express **16**, 8181-8189 (2008).
29. D.C. Hanna, J.E. Townsend, and A.C. Trooper, "Continuous-wave oscillation of holmium-doped silica fibre laser," Electron. Lett. **25**, 593-594 (1989).
30. M.E. Fermann, M.J. Andrejco, M.L. Stock, Y. Silberberg, and A.M. Weiner, "Passive mode locking in erbium fiber lasers with negative group delay," Appl. Phys. Lett. **62**, 910-912 (1993).
31. F. Krausz, M.E. Fermann, T. Brabec, P.F. Curley, M. Hofer, M.H. Ober, C. Spielmann, and A.J. Schmidt, "Femtosecond solid-state lasers," IEEE J. Quantum Electron. **28**, 2097-2122 (1992).
32. H.A. Haus, "Mode-locking of Lasers," IEEE J. Select. Topics Quantum Electron. **6**, 1173-1185 (2000).

## 1. Introduction

Ultrafast lasers operating in the 2 μm wavelength region are of great interest for a wide range of scientific and technological applications including fiber-optic and free-space communication, remote sensing and LIDAR, and various medical applications [1]. They are also of critical importance for nonlinear frequency conversion schemes into the mid-infrared and THz spectral region as they enable efficient pumping of non-oxide crystals without suffering from two-photon absorption [2-4]. Ho-doped gain materials are particularly attractive for high energy laser systems as their strong absorption band around 1.9 μm facilitates in-band pumping with commercial Tm:fiber lasers at a minimal quantum defect. Multi mJ-level Ho:YAG and Ho:YLF amplifier systems generating ps and sub-ps pulses have recently been demonstrated [5-7], but rely on complex seed sources such as optical parametric amplifiers or Raman-shifted fiber lasers. Alternative seed sources are Tm:fiber and Tm/Ho:fiber lasers [8-15] or frequency shifted Er:fiber lasers [16-23]. Previously demonstrated fiber-based seed laser either didn't cover the full gain spectrum of Ho-doped gain materials or lack flexibility and compactness. Ho:fiber lasers are the obvious choice for seeding Ho-doped gain materials. The transition between the $^5I_7$ and $^5I_8$ manifolds exhibits a large emission cross-section and a gain bandwidth suitable for the amplification of fs-pulses, which allows the direct generation of pulse-trains within the amplifer gain spectrum with variable parameters. However, most previously demonstrated Ho:fiber and solid state oscillators did not meet the spectral requirements for subsequent amplification [24-26].

We recently demonstrated a passively mode-locked Ho:fiber oscillator operating in the stretched-pulse regime which overcomes this limitation [27]. With proper dispersion management, the oscillator generated 160 fs pulses with 37 nm full-width-half-maximum (FWHM) spectrum around 2.06 µm simultaneously covering the wavelength range of both amplification bands of Ho:YLF amplifiers. We successfully used this oscillator to seed a high power Ho:YLF regenerative amplifier with $10^7$ gain [7]. The spectral bandwidth of this oscillator, however, exceeded the sub-10nm bandwidth of high-power Ho:YLF amplifiers by far, reducing the effectiveness of the seeding process.

In this paper, we demonstrate soliton operation of Ho:fiber oscillators in two different cavity configurations which can be fine-tuned and optimized to the seed requirements of Ho:YLF amplifiers. We demonstrate independent tuning of spectral bandwidth and center wavelength and present a relative intensity noise (RIN) characterization of the mode-locked pulse train.

## 2. Experimental setups

The schematics of the unidirectional ring oscillators are shown in Fig. 1. In both cavity configurations, the gain medium was a 1 m long Ho-doped double-clad fiber with a 10 µm-core, used in core-pumped configuration. The pump light from a commercial continuous wave Tm:fiber laser operating at 1950 nm was coupled into the cavity via a wavelength-division multiplexer. All cavity elements exhibit anomalous dispersion enabling soliton operation. In the first cavity configuration shown in Fig. 1(a), mode-locking was initiated and stabilized solely by nonlinear polarization evolution (NPE) whereas the rejection port of a polarizing beam splitter was used as the output port.

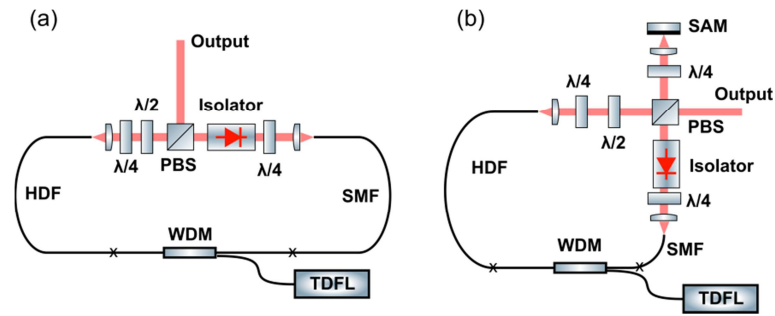

Fig. 1 (a) Schematic of the NPE mode-locked oscillator; (b) Schematic of the hybrid mode-locked oscillator. TDFL: Thulium-doped fiber laser; WDM: wavelength-division multiplexer; HDF: Holmium-doped fiber; λ/2: half-wave plate; λ/4: quarter-wave plate; PBS: polarizing beam splitter; SAM: saturable absorber mirror; SMF: single-mode fiber.

To further increase the self-starting capabilities, a hybrid mode-locked configuration shown in Fig. 1 (b) was set up with an additional commercial saturable absorber mirror (SAM). The SAM was implemented in a sigma-branch based on a polarization beam splitter and a quarter wave-plate. The SAM exhibited a modulation depth of 12 %, non-saturable losses of 8 % and a relaxation time of approximately 10 ps. The beam was focused down to a spot-size of 60 µm at $1/e^2$ by a 50 mm focal length $CaF_2$ lens. In the steady-state, the SAM was operated at only 15 % of the saturation fluence of 65 µJ cm$^{-2}$ given by the manufacturer. The small action of this additional saturable absorption in steady-state is typical for soliton operation of passively mode-locked laser oscillators. Here SAM mainly acts in the early stages of pulse formation where self q-switched operation typically occurs at repetition frequencies of about 100 kHz. We estimated the fluence on the SAM to > 1.3 mJ cm$^{-2}$ corresponding to a factor of > 20 above the saturation fluence. Tighter focusing was not possible due to damage of the SAM. By implementing this additional SAM we were able to reduce the mode-locking threshold by approximately 20 % compared to the oscillator mode-locked solely by NPE. This also resulted in a reduced sensitivity to alignment of the polarization optics used for NPE [28]. The main output characteristics such as pulse energy, pulse duration, spectral width and RIN performance were almost unaffected.

### 3. Output characteristics

Self-starting mode-locked operation was obtained in the cavity configurations solely mode-locked by NPE as well as with the additional SAM. The overall pulse energy was typically > 1 nJ with a maximum of 1.4 nJ. Owing to the large nonlinearity inside the cavity about 30 % of the pulse energy was coupled into the Kelly sidebands resulting in a typical pulse energy of 800 pJ for the central pulse. The optical-to-optical efficiencies were in the range of 7 – 15 %.

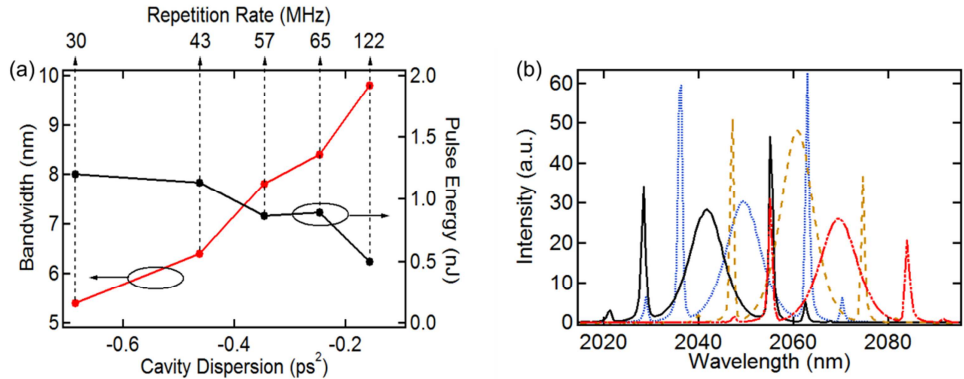

Fig. 2 (a) Spectral bandwidth at FWHM and pulse energy of the soliton pulse vs. cavity dispersion[AR1]. (b) tuning of the central wavelength -0.25 ps$^2$ cavity dispersion.

It is well known that the spectral bandwidth of a soliton laser can be tuned by changing the cavity dispersion according to $E_p = \frac{2|\beta_2|}{|\gamma|\tau}$, where $E_p$ is the pulse energy, $\beta_2$ the group velocity dispersion, $\gamma$ the nonlinear coefficient and $\tau$ the pulse duration. At lower cavity dispersions, shorter soliton pulses with broader bandwidth can be achieved. This behavior is illustrated in Fig. 2(a) for the first cavity configuration but also observable for the hybrid mode-locked laser. It can be seen that the reduction of the cavity dispersion from - 0.69 ps$^2$ to -0.16 ps$^2$ (by reducing SMF fiber length) results in an increase of the spectral bandwidth from 5 nm to 10 nm at FWHM. The cavity dispersion was calculated from the Kelly sideband spacing. As the cavity dispersion was changed via the length of the SMF section, the repetition rate consequently increased from 30 MHz to 122 MHz. The pulse energy in the soliton (with the energy in the Kelly sidebands subtracted) remained almost constant at around 800 pJ as can be seen from Fig. 2(a).

The broad gain bandwidth of Ho:fibers spanning 140 nm at FWHM [29] allows for tunable operation and hence gives the possibility to adapt the central wavelength to subsequent amplifiers. To demonstrate the tuning capability, the laser was operated at a net cavity dispersion of -0.25 ps$^2$ corresponding to a repetition frequency of 65 MHz. The central wavelength of the mode-locked pulse train could be tuned by adjusting the wave plates, and the free space to fiber coupling inside the cavity, utilizing the chromatic aberration of the coupling lenses. The drift in center wavelength during 3 days of operation was below our measurement limit of 0.7 nm. We achieved a tuning range from 2040 nm to 2070 nm as shown in Fig. 2(b)). A similar tuning behavior was observed over the full range of net cavity dispersions and repetition rates, respectively. Although the gain bandwidth of Ho:fibers would support an even larger tuning range, we were limited by the transmission bandwidths of the optical components used.

The tuning results highlight that this soliton oscillator can satisfy the seeding requirement of subsequent Ho:YLF amplifiers in terms of both central wavelength and spectral bandwidth. In order to optimize the oscillator for Ho:YLF regenerative amplifiers, we operated it at 65 MHz repetition rate and -0.25 ps$^2$ net cavity dispersion, respectively. The FWHM of the output spectrum was around 8 nm and the central spectral wavelength was carefully tuned to match the Ho:YLF gain spectrum. Fig. 3(a) shows the spectral overlap between the output of the oscillator and the single pass gain of Ho:YLF. The improvement in spectral energy density in the gain window of Ho:YLF amplifiers is more than one order of magnitude compared to our stretched-pulse approach reported in [27].

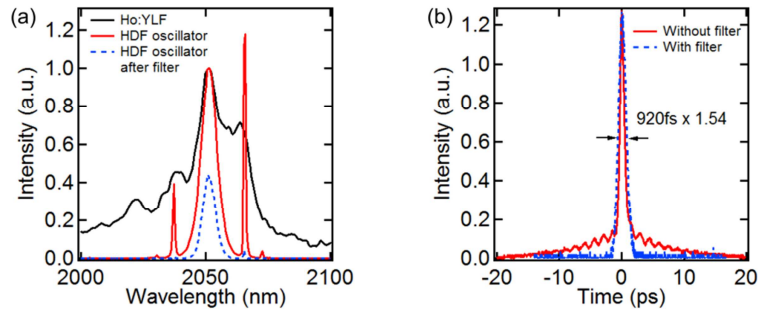

Fig. 3 (a) Spectral overlap between the output of the soliton oscillator with (red solid line) and without (blue dashed line) bandpass filter together with the single pass gain of Ho:YLF (black solid line). (b) Intensity autocorrelation trace of the soliton oscillator with (red solid line) and without (blue dashed line) the bandpass filter.

A potential drawback of soliton operation is the formation of strong Kelly sidebands and the resulting disturbances of the pulses. As shown in Fig. 3(a), in the chosen configuration, the Kelly sidebands are located outside the main gain window so they don't contribute to the pulse quality when seeding Ho:YLF amplifiers. To demonstrate the corresponding improvement in pulse quality, we suppressed the Kelly sidebands by filtering the main peak of the spectrum with a 10 nm bandpass filter and measured the second order autocorrelation shown in Fig. 3(b). As can be seen, filtering leads to a clean pulse shape with a FWHM of 920 fs. The Kelly sidebands mainly contribute to the strong pedestal spanning almost 40 ps. The corresponding time-bandwidth product of 0.47 is typical for soliton fiber lasers [30] and suggests that the observed solitons are slightly chirped. When assuming the Fourier-limited pulse duration of 560 fs and a pulse energy of 860 pJ (see Fig. 2 (a)), the central part of the generated pulses exhibit a soliton order of $N = 1.5$ close to $N = 1$ as expected from theory [31,32].

### 4. Relative intensity noise measurements

As amplitude noise is a crucial parameter for subsequent amplification, the relative intensity noise (RIN) of the mode-locked pulse train has been characterized. For the measurements, the laser output was sampled and the light was analyzed with a photodiode followed by a trans-impedance amplifier. In the spectral range from 1 Hz to 1 MHz, the intensity power spectral density was recorded using a vector signal analyzer with a 1.9 MHz low pass filter to prevent the vector analyzer being saturated by the high frequency portion. For the measurements from 1 MHz to 30 MHz we removed the low-pass filter and used an RF spectrum analyzer due to its larger dynamic range. Above a sidebands frequency of 1 MHz, the measurements were limited by the instrument noise floor.

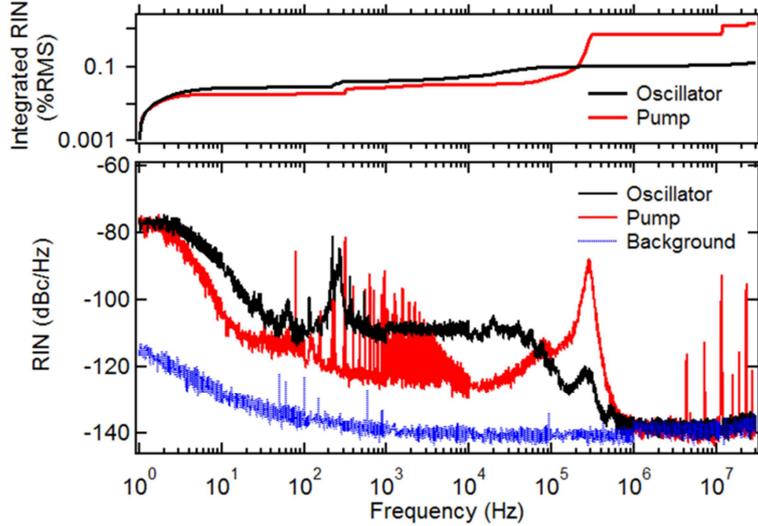

Fig. 4 RIN measurement of the Ho:fiber oscillator output after filter (black solid line) and the pump laser (red solid line) together with the instrument noise floor (blue dotted line).

Both the RIN of the pump laser and soliton oscillator output after filter were characterized. The corresponding RIN power spectral densities are shown in Fig 4. The low frequency noise contributions of all lasers mainly arise from temperature variation whereas the noise spikes from 10 Hz to 10 KHz have their origin in acoustics and vibrations. Additionally, the pump laser exhibits several sharp equidistant high-frequency contributions, which we attributed to transversal as well as longitudinal mode-beating.

The RIN integrated from 1 Hz to the Nyquist frequency of 32.5 MHz also shown in Fig. 4 was 1.43 % RMS and 0.13 % RMS for the pump laser and soliton oscillator. However, RIN is still dominated by the pump especially in low frequency region. Further reduction of the oscillator noise performance can therefore be expected by adopting a low-noise pump laser.

## 5. Conclusion

In conclusion, we demonstrated tunable soliton operation of a ps-Ho:fiber oscillator with independent tuning capability of spectral bandwidth from 5 nm to 10 nm and center wavelength from 2040 nm to 2070 nm while maintaining a pulse energy around 800 pJ. The oscillator was finely tuned and optimized to match the seeding requirement of Ho:YLF amplifiers. In particular we improved the power spectral density within the Ho:YLF gain bandwidth by more than one order of magnitude compared to a femtosecond oscillator [27]. These results, together with our previously demonstrated stretched-pulse operation [27], show that simple and compact passively mode-locked Ho:fiber lasers can be tailored in pulse durations from 0.2 to 1 ps, in central wavelengths from 2 µm to 2.1 µm and in spectral bandwidths from 5 nm to 100 nm. Ho:fiber oscillators are therefore a flexible platform which can be tailored for many future applications.